\documentclass[12pt,a4paper]{article}

\usepackage{authblk}
\usepackage{amsmath,amssymb}
\usepackage{graphicx}
\usepackage{amsfonts}
\usepackage{epsfig}

\usepackage{tikz}
\usepackage{pgflibrarysnakes}
\usetikzlibrary[snakes]

\usetikzlibrary{arrows,shapes,positioning}
\usetikzlibrary{decorations.markings}
\tikzstyle arrowstyle=[scale=1]
\tikzstyle directed=[postaction={decorate,decoration={markings,
		mark=at position .65 with {\arrow[arrowstyle]{stealth}}}}]
\tikzstyle reverse directed=[postaction={decorate,decoration={markings,
		mark=at position .65 with {\arrowreversed[arrowstyle]{stealth};}}}]

\renewcommand{\baselinestretch}{1.5}

\newcommand{\im}{\mathrm i}

\newcommand{\tr}{\operatorname{Tr}}

\newcommand{\ket}[1]{\left|#1\right\rangle}      
\newcommand{\bra}[1]{\left\langle #1\right|}     
\newcommand{\eq}{\begin{equation}}
\newcommand{\en}{\end{equation}}
\newcommand{\bear}{\begin{eqnarray}}
\newcommand{\ear}{\end{eqnarray}}

\title{Correlation functions of integrable $O(n)$ spin chains} 

\author{G.A.P. Ribeiro\footnote{E-mail: pavan@df.ufscar.br; On leave of absence from Departamento de F\'isica, Universidade Federal de S\~ao Carlos, PO Box 676, 13565-905, S\~ao Carlos-SP, Brazil. }}
\affil{School of Mathematics and Statistics, 
	University of Melbourne,\\ 
	Parkville, Victoria 3010, Australia}

\date{\today}

\begin{document}
\maketitle

\begin{abstract}
We study the correlation functions of the integrable $O(n)$ spin chain in the thermodynamic limit.
We addressed the problem of solving functional equations of the quantum Knizhnik Zamolodchikov type for density matrix related to the $O(n)$ spin chain. We give the explicit solution for two-sites density matrix elements
for the $O(n)$ which are then evaluated for the $n=3,4,\dots,8$ cases at zero temperature. 
\end{abstract}

\centerline{Keywords: Integrability, spin chains, correlation functions }

\thispagestyle{empty}
\newpage

\pagestyle{plain}
\pagenumbering{arabic}

\section{Introduction}

The correlation functions of integrable quantum model have been largely studied over the last decades \cite{BOOK-KBI,BOOK-JM}. The prominent case is the 
XXZ spin-$1/2$ chain, which was successfully studied in the thermodynamic limit by many different viewpoints, ranging from multiple integrals, algebraic Bethe ansatz methods, hidden Grassmann structure to functional equations of the quantum Knizhnik Zamolodchikov type \cite{TAKA,JMMN92,JM96,KMT00,GAS05,BOKO01,BOOS05,BOOS2,BGKS06,DGHK07,BJMST,SMIRNOV2009,KKMST09,SABGKT11,AuKl12}. There are also results for its higher-spin generalization\cite{BoWe94,Idzumi94,Kitanine01,DeMa10,GSS10,KNS013,RK2016,SMIRNOV19}, where explicit results were obtained by functional equations \cite{DeMa10,GSS10,KNS013,RK2016,SMIRNOV19}.

Nevertheless, the natural generalization of these correlation functions studies for high rank algebras $SU(n)$ for $n>2$ remained open for decades. This is mainly due to the intricate structure of the Bethe states, which are still being unveiled in the context of e.g algebraic Bethe ansatz\cite{SLAVNOV,WHEELER} and separation of variables\cite{NICOLI} with the aim of developing a manageable approach to evaluate correlations functions. However, very recently the first explicit results for short-distance correlations in the thermodynamic limit for the $SU(3)$ case were evaluated via the functional equations of quantum Knizhnik Zamolodchikov type \cite{BOOS18,RIBEIRO2018}. In addition, there are also recent results of the computation of current mean values  for the $SU(3)$ case \cite{POZSGAY}.

On the other hand, the high rank Lie algebras $B_n$, $C_n$ and $D_n$ \cite{RESHETIKHIN,KUNIBA} and superalgebras $OSp(n|2m)$ \cite{KULISH,MARTINS,MARTINS1997} are less studied. These cases have been treated via analytical Bethe ansatz \cite{RESHETIKHIN} and algebraic Bethe ansatz (nested Bethe ansatz) \cite{MARTINS,MARTINS1997} and its spectral and critical properties have been studied \cite{RESHETIKHIN,KLUMPER1990,MARTINSOSP}. Nevertheless, there are no explicit results for correlation functions. 
It is worth noting that it has recently appeared a general derivation of functional equations of the quantum Knizhnik-Zamolodchikov type in case of arbitrary simple Lie algebra \cite{RAZUMOV}. However, at finite temperature those equations comprise a system of two different functional equations similar to the approach used in \cite{BOOS18}, which therefore have a different structure from the single functional equation studied here that are obtained from an earlier approach \cite{AuKl12}.

In this work, we start the explicit calculation of correlation functions of integrable high rank spin chains (other than $SU(n)$) via the approach of functional equations of quantum Knizhnik Zamolodchikov type obtained from the framework devised in \cite{AuKl12}.
We aim to compute the reduced density matrix elements for finite segments for $O(n)$ spin chain in the thermodynamic limit, which would allow to evaluate the short-distance static correlation functions. Here, we start dealing with reduced density matrix for a system consisting of two neighboring sites within an infinite chain. 
We have used the functional equation for the $O(n)$ case and other properties like intertwining symmetry, asymptotic, analyticity and normalization to successfully compute the two-sites density matrix elements. Therefore, the main results of this paper are the  explicit solutions for the two-sites density matrix elements for the $O(n)$ spin chain, which was evaluated for the $n=3,4,\dots,8$ cases.

This paper is organized as follows. In section \ref{INTEGRA}, we outline the
integrable structure of the model. In section \ref{density}, we introduce the reduced density matrix and its
functional equations and symmetry properties. In the section \ref{2siteDM}, we present the zero
temperature solution for two-site correlation functions for the cases $O(3)$, $O(4)$ and $O(5)$. In the appendix we present additional results for the cases $O(6)$, $O(7)$ and $O(8)$. Our conclusions are given in section \ref{CONCLUSION}.

\section{The model}\label{INTEGRA}

The Hamiltonian of the integrable $O(n)$ spin chain can be written as \cite{RESHETIKHIN,KUNIBA,KULISH,MARTINS1997},
\eq
{\cal H}=-\sum_{i=1}^{L}\left(I_{i,i+1} -P_{i,i+1} + \frac{1}{\Delta} E_{i,i+1}\right),
\en
where  $L$ is number of sites, $\Delta=(n-2)/2$ and $I_{i,i+1}$, $P_{i,i+1}$ and $E_{i,i+1}$ are the identity, permutation and Temperley-Lieb operator acting on the sites $i$ and $i+1$. 
For completeness we list their matrix elements $(I_{i,i+1})_{ac}^{bd}=\delta_{a,b}\delta_{c,d}$, $(P_{i,i+1})_{ac}^{bd}=\delta_{a,d}\delta_{b,c}$ and $(E_{i,i+1})_{ac}^{bd}=\delta_{a,n+1-c}\delta_{b,n+1-d}$ for $1\leq a,b,c,d \leq n$. The Hilbert space is $V^{\otimes L}$, where $V=\mathbb{C}^n$.

Here, we recall that the Hamiltonian ${\cal H}$ is obtained from the logarithmic
derivative of the row-to-row transfer matrix 
$T(\lambda)=\tr_{\cal A}{[ R_{{\cal A},L}(\lambda)\dots
R_{{\cal A}, 1}(\lambda)]}$, such that ${\cal H}=\frac{d}{d\lambda}\log{T(\lambda)}\Big|_{\lambda=0} 
= \sum_{i=1}^L h_{i,i+1}$, where $h_{i,i+1}=P_{i,i+1}\frac{d}{d\lambda} R_{i,i+1}\Big|_{\lambda=0}$ and that  
the $R$-matrix is a solution the well-known Yang-Baxter equation
\eq
R_{12}(\lambda-\mu) R_{13} (\lambda) R_{23} (\mu) =R_{23}(\mu) R_{13}(\lambda)  R_{12} (\lambda-\mu).
\label{yang-baxter}
\en
For the case of $O(n)$ integrable chain, the $R$-matrix can be conveniently written as \cite{RESHETIKHIN,KUNIBA,KULISH,MARTINS1997}
\eq
R_{12}(\lambda)= \frac{\lambda}{\lambda+1} I_{12} + \frac{1}{\lambda+1} P_{12} -\frac{\lambda }{(\lambda+1)(\lambda+\Delta)}E_{12}.
\label{Rmatrix}
\en

This matrix has the important properties of regularity, unitarity and crossing as given below,
\bear
 R_{12}(0)&=&P_{12}, \\
 R_{12} (\lambda) R_{21} (-\lambda) &=& I_{12}, \\
 R_{12} (\lambda) &=& \varrho(\lambda)(V\otimes I) R_{12}^{t_2} (-\lambda-\rho) (V\otimes I),
\ear
where $t_2$ is transposition in the second space, the crossing parameter is given by $\rho= \Delta$, the crossing matrix $V$ is a unity anti-diagonal matrix $V=\mbox{anti-diagonal}(1,1,\dots,1)$ and $\varrho(\lambda)=-\lambda(1-\lambda-\Delta)/((\lambda+1)(\lambda+\Delta))$.

\section{Density matrix and functional equations}\label{density}

In \cite{GoKlSe04} it was 
developed a scheme to deal with thermal correlation functions of
integrable spin chain, which was later used to deal with 
the case of higher-spin integrable $SU(2)$ chains \cite{GSS10,KNS013,RK2016} and also more recently to the case of integrable $SU(n)$ spin chains \cite{BOOS18,RIBEIRO2018}.

\begin{figure}[h]
	\begin{center}
		\begin{tikzpicture}[scale=1.65]
		\draw (0,0) [->,color=black, thick, rounded corners=7pt] +(0.8,1.75)--(0.8,3);
		\draw (0,0) [-,color=black, thick, rounded corners=7pt] +(0.8,0)--(0.8,1.25);
		
		\draw (0,0) [->,color=black, thick, rounded corners=7pt] +(1.4,1.75)--(1.4,3);
		\draw (0,0) [-,color=black, thick, rounded corners=7pt] +(1.4,0)--(1.4,1.25);
		
		\draw (0,0) [->,color=black, thick, rounded corners=7pt] +(2.05,1.75)--(2.05,3);
		\draw (0,0) [-,color=black, thick, rounded corners=7pt] +(2.05,0)--(2.05,1.25);
		
		\draw (0.8,-0.25) node {$\lambda_1$};
		\draw (1.4,-0.25) node {$\lambda_2$};
		\draw (1.75	,-0.25) node {$\dots$};
		\draw (2.05,-0.25) node {$\lambda_m$};
		
		\draw (0.8,1.25)[fill=black]  circle (0.15ex);
		\draw (0.8,1.75)[fill=black]  circle (0.15ex);
		
		\draw (1.4,1.25)[fill=black]  circle (0.15ex);
		\draw (1.4,1.75)[fill=black]  circle (0.15ex);
		
		\draw (2.05,1.25)[fill=black]  circle (0.15ex);
		\draw (2.05,1.75)[fill=black]  circle (0.15ex);
		
		\draw (-0.25,0) [->,color=black, thick, rounded corners=7pt] +(0.5,0.5)--(3.,0.5);
		\draw (-0.25,0) [->,color=black, thick, rounded corners=7pt] +(0.5,1.0)--(3.,1);
		
		\draw (-0.25,0) [->,color=black, thick, rounded corners=7pt] +(0.5,2.5)--(3.,2.5);
		\draw (-0.25,0) [->,color=black, thick, rounded corners=7pt] +(0.5,2.0)--(3.,2.0);

		\draw (-1.5,1.5) node {$\widetilde{D}_m(\lambda_1,\lambda_2,\dots,\lambda_m)=$};
		
		\draw (2.45,0.6) node {$u_1$};
		\draw (2.45,1.1) node {$u_2$};
		\draw (2.45,1.6) node {$\vdots$};
		\draw (2.45,2.6) node {$u_{N}$};
		\draw (2.45,2.1) node {$u_{N-1}$};
		
		\draw (0,1.51) node {$\Phi_0$};
		\draw (0,0) [-,color=black, thick, rounded corners=7pt] +(0.25,0)--(0.25,3)--(-0.3,1.5)--(0.25,0);
		\draw (3.25,1.51) node {$\Phi_0$};
		\draw (0,0) [-,color=black, thick, rounded corners=7pt] +(3.,0)--(3.,3)--(3.55,1.5)--(3.,0);
		\end{tikzpicture}
	\end{center}
	\caption{Graphical illustration of the un-normalized density matrix
		$\widetilde{D}_m(\lambda_1,\lambda_2,\dots,\lambda_m)= \bra{\Phi_0}{\cal T}_1(\lambda_1) {\cal T}_2(\lambda_2)\cdots {\cal T}_m(\lambda_m) \ket{\Phi_0}$. The infinitely many column-to-column
		transfer matrices to the left and to the right are replaced by the boundary
		states they project onto.}
	\label{fig-dmatrix}
\end{figure}
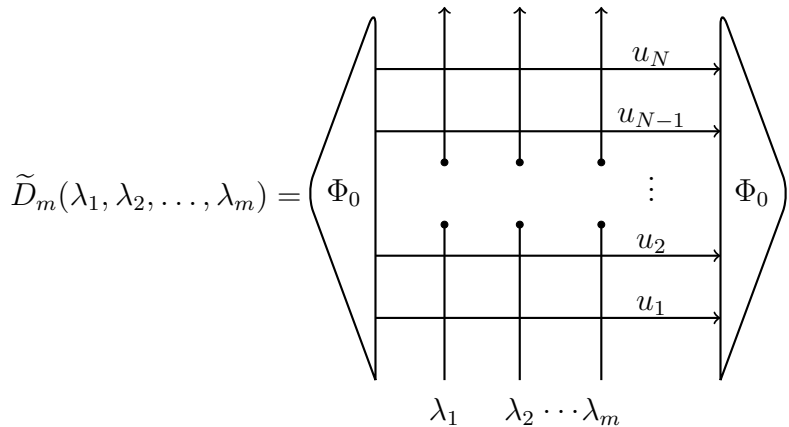

Within this approach the main object of concern is inhomogeneous reduced density matrix defined in the
thermodynamic limit $L \to \infty$ and  finite Trotter number $N$ \cite{GoKlSe04} (see Figure \ref{fig-dmatrix}),
\eq
D_m(\lambda_1,\lambda_2,\cdots,\lambda_m)=\frac{  \bra{\Phi_0}{\cal T}_1(\lambda_1) {\cal T}_2(\lambda_2)\cdots {\cal T}_m(\lambda_m) \ket{\Phi_0}}{ \Lambda_0(\lambda_1)\Lambda_0(\lambda_2)\cdots \Lambda_0(\lambda_m) },
\label{i-densitymatrix}
\en
where ${\cal T}_j(x)$ is the monodromy matrix  ${\cal
T}_j(x)=R_{j,N}(x-u_N)\dots R_{j,2}(x-u_2) R_{j,1}(x-u_1)$  associated to the quantum transfer matrix $t_j^{QTM}(x)=\tr[{\cal T}_j(x)]$ and $\ket{\Phi_0}$ is the eigenstate associated to the leading eigenvalue  $\Lambda_0(x)$ of the quantum transfer matrix.

It is worth to note that the connection with the physical density matrix for $m$-sites is obtained from the 
inhomogeneous reduced density matrix via the homogeneous limit $\lambda_j \to 0$ and the Trotter limit $N\to\infty$,
\eq
\check{D}_{[1,m]}=\lim_{N\rightarrow\infty} \lim_{\lambda_1,\cdots,\lambda_m\rightarrow 0} D_m(\lambda_1,\lambda_2,\cdots,\lambda_m).
\en

This poses the question of how to efficiently compute the referred inhomogeneous reduced density so that these limit can be taken. 
In the standard case of $SU(2)$ density matrix, this is e.g addressed by the derivation of a set of discrete functional equations by use of the integrability structure plus the crossing symmetry \cite{AuKl12}. The solution of these functional equations provides the complete determination of the reduced density matrix at finite temperature and zero temperature. 

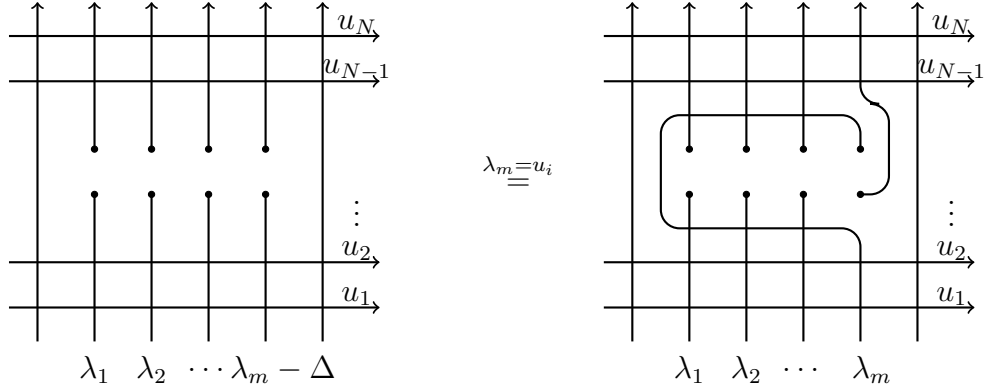
\begin{figure}
	\begin{center}
		\begin{minipage}{0.5\linewidth}
		\begin{tikzpicture}[scale=1.5]
		\draw (0,0) [->,color=black, thick, rounded corners=7pt] +(0.5,0.)--(0.5,3);
		
		\draw (0,0) [->,color=black, thick, rounded corners=7pt] +(1.,1.7)--(1.,3);
		\draw (0,0) [-,color=black, thick, rounded corners=7pt] +(1.,0)--(1.,1.3);

		\draw (0,0) [->,color=black, thick, rounded corners=7pt] +(1.5,1.7)--(1.5,3);
		\draw (0,0) [-,color=black, thick, rounded corners=7pt] +(1.5,0)--(1.5,1.3);
		
		\draw (0,0) [->,color=black, thick, rounded corners=7pt] +(2.,1.7)--(2.,3);
		\draw (0,0) [-,color=black, thick, rounded corners=7pt] +(2.,0)--(2.,1.3);

		\draw (0,0) [->,color=black, thick, rounded corners=7pt] +(2.5,1.7)--(2.5,3);
		\draw (0,0) [-,color=black, thick, rounded corners=7pt] +(2.5,0)--(2.5,1.3);

		\draw (0,0) [->,color=black, thick, rounded corners=7pt] +(3.,0.)--(3.,3);
		
		\draw (1.,-0.25) node {$\lambda_1$};
		\draw (1.5,-0.25) node {$\lambda_2$};
		\draw (2.	,-0.25) node {$\dots$};
		\draw (2.65,-0.25) node {$\lambda_m- \Delta$};
				
		\draw (1.,1.3)[fill=black]  circle (0.15ex);
		\draw (1.,1.7)[fill=black]  circle (0.15ex);

		\draw (1.5,1.3)[fill=black]  circle (0.15ex);
		\draw (1.5,1.7)[fill=black]  circle (0.15ex);
		
		\draw (2.,1.3)[fill=black]  circle (0.15ex);
		\draw (2.,1.7)[fill=black]  circle (0.15ex);

		\draw (2.5,1.3)[fill=black]  circle (0.15ex);
		\draw (2.5,1.7)[fill=black]  circle (0.15ex);

		\draw (-0.25,0) [->,color=black, thick, rounded corners=7pt] +(0.5,0.3)--(3.5,0.3);
		\draw (-0.25,0) [->,color=black, thick, rounded corners=7pt] +(0.5,0.7)--(3.5,0.7);
		
		\draw (-0.25,0) [->,color=black, thick, rounded corners=7pt] +(0.5,2.7)--(3.5,2.7);
		\draw (-0.25,0) [->,color=black, thick, rounded corners=7pt] +(0.5,2.3)--(3.5,2.3);
		
		\draw (3.3,0.4) node {$u_1$};
		\draw (3.3,0.8) node {$u_2$};
		\draw (3.3,1.2) node {$\vdots$};
		\draw (3.3,2.8) node {$u_{N}$};
		\draw (3.3,2.4) node {$u_{N-1}$};

		\end{tikzpicture}
		\end{minipage}%
		\begin{minipage}{0.5\linewidth}
			\begin{tikzpicture}[scale=1.5]
		\draw (0,0) [->,color=black, thick, rounded corners=7pt] +(0.5,0.)--(0.5,3);
		
		\draw (0,0) [->,color=black, thick, rounded corners=7pt] +(1.,1.7)--(1.,3);
		\draw (0,0) [-,color=black, thick, rounded corners=7pt] +(1.,0)--(1.,1.3);
		
		\draw (0,0) [->,color=black, thick, rounded corners=7pt] +(1.5,1.7)--(1.5,3);
		\draw (0,0) [-,color=black, thick, rounded corners=7pt] +(1.5,0)--(1.5,1.3);
		
		\draw (0,0) [->,color=black, thick, rounded corners=7pt] +(2.,1.7)--(2.,3);
		\draw (0,0) [-,color=black, thick, rounded corners=7pt] +(2.,0)--(2.,1.3);
		
		\draw (0,0) [->,color=black, thick, rounded corners=7pt] +(2.5,1.3)--(2.75,1.3)--(2.75,2.1)--(2.5,2.1)--(2.5,3.0);
		\draw (0,0) [-,color=black, thick, rounded corners=7pt] +(2.5,0)--(2.5,1.)--(0.75,1.) --(0.75,2.)--(2.5,2.)--(2.5,1.7);
		
		\draw (0,0) [->,color=black, thick, rounded corners=7pt] +(3.,0.)--(3.,3);
		
		\draw (1.,-0.25) node {$\lambda_1$};
		\draw (1.5,-0.25) node {$\lambda_2$};
		\draw (2.	,-0.25) node {$\dots$};
		\draw (2.6,-0.25) node {$\lambda_m $};
		
		\draw (1.,1.3)[fill=black]  circle (0.15ex);
		\draw (1.,1.7)[fill=black]  circle (0.15ex);
		
		\draw (1.5,1.3)[fill=black]  circle (0.15ex);
		\draw (1.5,1.7)[fill=black]  circle (0.15ex);
		
		\draw (2.,1.3)[fill=black]  circle (0.15ex);
		\draw (2.,1.7)[fill=black]  circle (0.15ex);
		
		\draw (2.5,1.3)[fill=black]  circle (0.15ex);
		\draw (2.5,1.7)[fill=black]  circle (0.15ex);
		
		\draw (-0.25,0) [->,color=black, thick, rounded corners=7pt] +(0.5,0.3)--(3.5,0.3);
		\draw (-0.25,0) [->,color=black, thick, rounded corners=7pt] +(0.5,0.7)--(3.5,0.7);
		
		\draw (-0.25,0) [->,color=black, thick, rounded corners=7pt] +(0.5,2.7)--(3.5,2.7);
		\draw (-0.25,0) [->,color=black, thick, rounded corners=7pt] +(0.5,2.3)--(3.5,2.3);
		
		\draw (3.3,0.4) node {$u_1$};
		\draw (3.3,0.8) node {$u_2$};
		\draw (3.3,1.2) node {$\vdots$};
		\draw (3.3,2.8) node {$u_{N}$};
		\draw (3.3,2.4) node {$u_{N-1}$};

			\draw (-0.5,1.5) node {$\stackrel{\lambda_m=u_i}{=}$};
		
			\end{tikzpicture}
		\end{minipage}
	\end{center}
\caption{Graphical illustration of the functional equation (\ref{qKZ}).}
\label{pic-qKZ}
\end{figure}

Similarly for the case of integrable $O(n)$ chains, there exist the necessary integrable structure and the crossing symmetry, which allow us to proceed along the same lines as \cite{AuKl12} in deriving the discrete functional equation for the density matrix. In this case, the equation reads,
\eq
D_m(\lambda_1,\lambda_2,\cdots,\lambda_m- \Delta)=A_m(\lambda_1,\cdots,\lambda_m)[D_m(\lambda_1,\cdots,\lambda_m)],
\label{qKZ}
\en
where the linear operator $A_m$ can be written as,
\bear
A_m(\lambda_1,\cdots,\lambda_m)[B]&:=&\tr_{m}\Big[R_{1,2}(-\lambda_{1,m}) \cdots R_{m-1,m}(-\lambda_{m-1,m})(P_{s})_{m,m+1} \nonumber \\
&\times &(B\otimes I_{m+1}) R_{m-1,m}(\lambda_{m-1,m})\cdots R_{1,2}(\lambda_{1,m})\Big],
\label{A-qKZ}
\ear
where $\lambda_{i,j}=\lambda_i-\lambda_j$ and $P_{s}$ 
is a (not normalized) projector onto the two-site singlet and the partial trace is taken in the $m$-th vertical space. The graphical depiction of the functional equation (\ref{qKZ}) is given in Figure \ref{pic-qKZ}.

Moreover the analyticity properties of the density matrix are clear, since its matrix elements can be written as,
\eq
{D_m}(\lambda_1,\cdots,\lambda_m)^{\epsilon_1,\dots,\epsilon_m}_{\bar\epsilon_1,\dots,\bar\epsilon_m}=\frac{  
	Q(\lambda_1, \dots, \lambda_m)}{ \Lambda_0(\lambda_1)\cdots \Lambda_0(\lambda_m) },
\label{analyticity-densitymatrix}
\en
where $Q(\lambda_1,\dots,\lambda_m)$ is a multivariate polynomial of degree up to $2N$ in each variable (for suitable normalization of the $R$-matrix), as is the case in the spin-$1$ $SU(2)$ chain \cite{KNS013}.

Besides, we have the properties of normalization $\tr [D_m(\lambda_1,\cdots,\lambda_2)]=1$, asymptotic condition,
\eq
\lim_{\lambda_m\rightarrow\infty} D_m(\lambda_1,\cdots,\lambda_m)=D_{m-1}(\lambda_1,\cdots,\lambda_{m-1})\otimes \mbox{Id},
\label{asymp}
\en
and intertwining symmetry relations
\bear
(R_{k,k+1}(\lambda_k-\lambda_{k+1}))^{-1}&D_m(\lambda_1,\cdots,\lambda_k,\lambda_{k+1},\cdots,\lambda_m)&R_{k,k+1}(\lambda_k-\lambda_{k+1})=  \nonumber\\ 
=& D_m(\lambda_1,\cdots,\lambda_{k+1},\lambda_{k},\cdots,\lambda_m)&,
\label{intertwin}
\ear
which altogether resolves the under-determinacy of the functional equations. This means that with the use of the functional equation (\ref{qKZ}), analyticity (\ref{analyticity-densitymatrix}), normalization, asymptotic (\ref{asymp}) and intertwining (\ref{intertwin}) properties we can fully determine the density matrix,
as we verified for characteristic cases for finite Trotter number $N$.

In the next section we are going to illustrate the feasibility of this approach by the computation of the two-sites ($m=2$) density matrix $D_2(\lambda_1,\lambda_2)$. It is worth to emphasize  that  the two-site density matrix for the case of integrable $O(n)$ spin chains is already a non-trivial result in the homogeneous limit,
similarly to the case of higher-spin $SU(2)$ chains \cite{KNS013,RK2016}. This is a consequence of the fact that we need three independent relations to determine the three unknown coefficients $\rho_{i}(0,0)$, $i=1,2,3$. Therefore the normalization and the knowledge of ground state energy are not sufficient conditions and the functional equations provides precisely the additional constraint.

\section{Computation of the two-site density matrix}\label{2siteDM}

Due to the $O(n)$ invariance, the two-sites reduced density matrix can be
written as a combination the identity, permutation and Temperley-Lieb operator, likewise the $R$-matrix (\ref{Rmatrix}), therefore we have that
\eq
D_2(\lambda_1,\lambda_2)=\rho_{1}(\lambda_1,\lambda_2) I_{12} + \rho_{2}(\lambda_1,\lambda_2) P_{12} +\rho_{3}(\lambda_1,\lambda_2) E_{12},
\label{D2rep}
\en
which leaves us with the unknown functions $\rho_i(\lambda_1,\lambda_2)$ to be determined. 

Therefore, one has to work out the functional equations for the expansion coefficients $\rho_i(\lambda_1,\lambda_2)$. This is done by replacing the above expression for the density matrix (\ref{D2rep}) into the functional equation (\ref{qKZ}), which results into the following set of functional equations,
\eq
\left(\begin{array}{c}
	\rho_{1}(\lambda_1-\Delta,\lambda_2) \\
	\rho_{2}(\lambda_1-\Delta,\lambda_2) \\
	\rho_{3}(\lambda_1-\Delta,\lambda_2)
\end{array}\right)=
{\cal A}(\lambda) \cdot\left(\begin{array}{c}
	\rho_{1}(\lambda_1,\lambda_2) \\
	\rho_{2}(\lambda_1,\lambda_2) \\
	\rho_{3}(\lambda_1,\lambda_2)
\end{array}\right), \qquad \lambda_1=u_i,
\label{matrixL-qKZ}
\en
where $\lambda=\lambda_1-\lambda_2$ and the matrix ${\cal A}(\lambda)$ is given by,
\eq
{\cal A}(\lambda)=\left(\begin{array}{ccc}
\frac{\lambda^2(\lambda^2-(\Delta^2+1))}{(\lambda^2-1)(\lambda^2 -\Delta^2)} & -\frac{\lambda(\Delta^2 + \lambda - \lambda^2)}{(\lambda^2-1)(\lambda^2-\Delta^2)} & \frac{\lambda}{(\lambda+1)(\lambda+\Delta)} \\ 
\frac{2\Delta\lambda^2}{( \lambda^2-1)(\lambda^2-\Delta^2 )} & \frac{\lambda(\Delta + \lambda + \lambda \Delta - \lambda^2)}{(\lambda^2-1)(\lambda^2 -\Delta^2)} & 	\frac{\lambda( \lambda-\Delta )}{(\lambda+1)(\lambda+\Delta)} \\ 
\frac{2\Delta((1 + \Delta)\Delta - \lambda^2)}{( \lambda^2-1)( \lambda^2-\Delta^2)} & \frac{\Delta^2 + \Delta(-1 + 2\Delta(1 + \Delta))\lambda - (-1 + \Delta)\Delta\lambda^2 - (1 + 2\Delta)\lambda^3 + \lambda^4}{(\lambda^2-1)( \lambda^2-\Delta^2)} & \frac{\Delta - \lambda}{\Delta + (1 + \Delta)\lambda + \lambda^2}
\end{array}\right).
\nonumber
\en

For convenience, we define the intermediate auxiliary functions
$\Omega_0(\lambda_1,\lambda_2)=1= \tr[D_2(\lambda_1,\lambda_2)]$, $\Omega_1(\lambda_1,\lambda_2)= \tr[P_{12} D_2(\lambda_1,\lambda_2)]$ and $\Omega_2(\lambda_1,\lambda_2)=\tr[E_{12} D_2(\lambda_1,\lambda_2)]$ such that,
 
\eq
\left(\begin{array}{c}
	1 \\
	\Omega_{1}(\lambda_1,\lambda_2) \\
	\Omega_{2}(\lambda_1,\lambda_2)
\end{array}\right)=
\left(\begin{array}{ccc} 
	n^2 & n & n \\
	n & n^2 & n \\
	n & n & n^2  
	\end{array}\right)\cdot
  \left(\begin{array}{c}
	\rho_{1}(\lambda_1,\lambda_2) \\
	\rho_{2}(\lambda_1,\lambda_2) \\
	\rho_{3}(\lambda_1,\lambda_2)
\end{array}\right), 
\label{aux-func}
\en
where it is worth to recall that $\Delta=(n-2)/2$ (or conversely $n=2(\Delta+1)$).

Therefore, we can chose suitable new functions
\bear
1&=&\Omega_0(\lambda_1,\lambda_2), \nonumber \\
\omega_1(\lambda_1,\lambda_2)&=&\frac{1}{\lambda^2-1}\left[ 1 - \Omega_1(\lambda_1,\lambda_2) + \frac{(\lambda^2-\Delta)}{(\lambda^2-\Delta^2)}\Omega_2(\lambda_1,\lambda_2)\right],  \label{aux-func2}
\\
\omega_2(\lambda_1,\lambda_2)&=&\frac{ \lambda}{(\lambda+1)(\lambda^2-\Delta^2)}  \Omega_2(\lambda_1,\lambda_2), \nonumber
\ear
which can almost completely decouple the set of equations.

The set of equation (\ref{matrixL-qKZ}) in terms of the new functions are written as,
\eq
\left(\begin{array}{c}
	1 \\
	\omega_{1}(\lambda_1-\Delta,\lambda_2) \\
	\widetilde\omega_{2}(\lambda_1-\Delta,\lambda_2)
\end{array}\right)=
\left(\begin{array}{ccc} 
	1 & 0 & 0 \\
	\frac{1}{\lambda-1}-\frac{1}{\lambda}+\frac{1}{\lambda-\Delta}-\frac{1}{\lambda -\Delta+1} & -1 & 0 \\
	\alpha(\lambda)(\frac{1}{\lambda-1}-\frac{1}{\lambda})  & -\alpha(\lambda) & 1  	
\end{array}\right)\cdot
\left(\begin{array}{c}
	1 \\
	\omega_{1}(\lambda_1,\lambda_2) \\
	\widetilde\omega_{2}(\lambda_1,\lambda_2)
\end{array}\right), 
\label{set-func}
\en
for $\lambda_1=u_i$ and where we conveniently introduce the function  $\widetilde{\omega}_2(\lambda_1,\lambda_2) =\alpha(\lambda) \omega_2(\lambda_1,\lambda_2)$ and $\alpha(\lambda)=\frac{\Gamma\left(\frac{\lambda-1}{\Delta}\right)}{\Gamma\left(\frac{\lambda+1}{\Delta}\right)}$ such that $\frac{\alpha(\lambda)}{\alpha(\lambda-\Delta)}=\frac{(\lambda-\Delta-1)}{(\lambda-\Delta+1)}$.
 
Besides, one has to impose the intertwining symmetry which implies that the density matrix is symmetric $D_2(\lambda_1,\lambda_2)=D_2(\lambda_2,\lambda_1)$ under the exchange of the arguments and therefore the same applies to  its expansion coefficients $\rho_i(\lambda_1,\lambda_2)$.

In addition, the asymptotic condition implies that 
\bear
\lim_{\lambda_j\rightarrow\infty} \rho_k(\lambda_1,\lambda_2)=\begin{cases}
\frac{1}{n^2}, \qquad k=1  \\
0, \qquad k \neq 1
\end{cases}
\ear
which implies that $\lim_{\lambda_j\rightarrow\infty}\omega_k(\lambda_1,\lambda_2)=0$.

\subsection{Zero temperature solution}

At zero temperature, the functional equations (\ref{set-func}) hold for arbitrary values
$\lambda_1$. This is due to the fact that at zero temperature one has to take the Trotter
limit ($N\to\infty$) and therefore the horizontal spectral parameters $u_i$ can take an infinite number of continuous values. Besides, we assume that at zero temperature the $\omega_i$ functions depend on the difference of the arguments, as in \cite{SABGKT11,KNS013,RK2016,SMIRNOV19,RIBEIRO2018}. The equation for $\omega_{1}(\lambda_1,\lambda_2)=\omega_1(\lambda)$ is fully decoupled as given by,
\eq
\omega_1(\lambda-\Delta)+\omega_1(\lambda)=\frac{1}{\lambda-1}-\frac{1}{\lambda}+\frac{1}{\lambda-\Delta}-\frac{1}{\lambda -\Delta+1}, 
\label{eq1}
\en
which can be solved via Fourier transform resulting in,
\bear
\omega_1(\lambda)=-\frac{d}{d\lambda} \log\left[ \frac{\Gamma(1+\frac{\lambda}{2\Delta}) \Gamma(\frac{1}{2}-\frac{\lambda}{2\Delta}) \Gamma(\frac{1}{2}+\frac{1}{2 \Delta} +  \frac{\lambda}{2\Delta}) \Gamma(\frac{1}{2\Delta}-  \frac{\lambda}{2\Delta})}{ \Gamma(1- \frac{\lambda}{2\Delta}) \Gamma(\frac{1}{2}+ \frac{\lambda}{2\Delta}) \Gamma(\frac{1}{2} +\frac{1}{2\Delta} -  \frac{\lambda}{2\Delta}) \Gamma(\frac{1}{2\Delta}+   \frac{\lambda}{2\Delta})}\right],
\label{sol-lam}
\ear
where $\Gamma(z)$ is the gamma function.

Taking the homogeneous limit $\lambda_i=\lambda=0$, result in the ground state energy of the $O(n)$ spin chain \cite{RESHETIKHIN,MARTINSOSP},
\eq
E_{gs}=\omega_1(0)=-\frac{1}{\Delta}\left[ 2 \log(2) -\psi_0\left(\frac{1}{2\Delta}\right)+ \psi_0\left(\frac{1}{2}+\frac{1}{2\Delta}\right) \right],
\en
where $\psi_0(z)$ is the digamma function $\psi_0(z)=\frac{d}{dz}\log \Gamma(z)$.

On the one hand the equation (\ref{eq1}) for $\omega_1(\lambda)$  is an inhomogeneous equation, on the other hand its inhomogeneity is a rational function, which allows for closed analytically solution (\ref{sol-lam}). Nevertheless, the equation for the $\widetilde\omega_2(\lambda)$ is also an inhomogeneous equation given as,
\eq
\widetilde\omega_2(\lambda-\Delta)-\widetilde\omega_2(\lambda)=	\alpha(\lambda)\left[\frac{1}{\lambda-1}-\frac{1}{\lambda}  -\omega_1(\lambda)\right], 
\label{eq2} 	
\en
however, in this case the functions $\alpha(\lambda)$ and $\omega_1(\lambda)$ (\ref{sol-lam}) appear in the inhomogeneous function. Therefore, this equation is in general very hard to solve analytically. The exception is the case $n=3$ (or conversely $\Delta=1/2$), where both functions $\alpha(\lambda)$ and $\omega_1(\lambda)$ becomes rational functions themselves, which allows for an analytical solution.

In general, we can write an integral expression as the solution. In order to do so, we use analyticity in the variable $\lambda$ and Fourier
transform the above equations. The resulting equations are algebraically
solved for the Fourier coefficients and yield product expressions. Then, we
Fourier transform back and find integrals of convolution type

\bear
\widetilde\omega_2(\lambda) &=& \int_{-\infty}^{\infty}  K(\lambda-\mu) \varphi(\mu) \frac{d\mu}{2\pi}, 
\label{gconv}
\ear
where 
\eq
K(z)= \int_{{\mathbb R} +\im 0} \frac{ e^{\im k z}}{1-e^{-\Delta k}} dk,
\en 
and 
\bear
\varphi(\lambda)&=& \alpha(\lambda)\left[ \frac{1}{\lambda- 1}-\frac{1}{\lambda}  -\omega_1( \lambda)\right].
\ear
The integral expression can be evaluated numerically. In what follows we show the results for the cases $n=3,4,5$ and the cases $n=6,7,8$ are presented in the appendix.

\subsubsection{The $O(3)$ case}

We have evaluated numerically the convolution integral (\ref{gconv}) at the homogeneous point $\lambda=0$ and obtained the value of $\widetilde\omega_2(0)$. In Table 1, we show the results from exact diagonalization for the ground state energy $\omega_1(0)$ and the correlation function $\widetilde{\omega}_2(0)$. The numerical results for finite lattices show agreement with the infinite lattice result obtained from the solution of the functional equations. 

\begin{table}[h]
	\begin{center}
		\begin{tabular}{|l|l|l|}
			\hline  
			Length & $\omega_{1}(0)$  & $\widetilde	\omega_{2}(0)$  \\ 
			\hline
			$L=2$ & $-6.000000000000000$ &  $-3.000000000000000$ \\
			\hline
			$L=4$ & $-4.350781059358212$ &  $-1.937042571331636$ \\
			\hline
			$L=6$ & $-4.146234978548967$ &  $-1.809210082780898$ \\
			\hline
			$L\rightarrow\infty$ & $-4.000000000000000$ & $-1.719824178261938
			$ \\
			\hline 
		\end{tabular}
	\end{center}
	\caption{Comparison of numerical results from exact diagonalization for $L=2,4,6$ sites with the analytical result in the thermodynamic limit for correlation $\omega_i(0)$ for the $O(3)$ case.}
\end{table}

Having the results for $\omega_i(0)$, we can use the relations (\ref{aux-func}-\ref{aux-func2}) to evaluate the expansion coefficient of the density matrix (see Table 2). We also show the results obtained from the exact diagonalization. 

\begin{table}[h]
	\begin{center}
		\begin{tabular}{|l|l|l|l|}
			\hline  
			Length & $\rho_{1}(0,0)$  & $\rho_{2}(0,0)$ & $\rho_{3}(0,0)$ \\ 
			\hline
			$L=2$ & $0.000000000000000$ &  $\hspace{0.35cm}0.000000000000000$ & $0.33333333333333$ \\
			\hline
			$L=4$ & $0.051321778178776$ &  $-0.028127541270379$ & $0.20749554006738$  \\
			\hline
			$L=6$ & $0.057286824340209$  & $-0.030682311157652$  & $0.19215517147035$ \\
			\hline
			$L\rightarrow\infty$ &  $0.061350915507139 $ & $-0.032041025072214$  & $0.18132161188412$  \\
			\hline 
		\end{tabular}
	\end{center}
	\caption{Comparison of numerical results from exact diagonalization for $L=2,4,6$ sites with the analytical result in the thermodynamic limit for the $\rho_i(0,0)$   for the $O(3)$ case.}
\end{table}

As previously mentioned, the case $n=3$  can be analytically solved since both functions $\alpha(\lambda)$ and $\omega_1(\lambda)$ becomes rational functions themselves,
\bear
\alpha(\lambda)&=&\frac{1}{2\lambda(2\lambda+1)(2\lambda-1)(2\lambda-2)}, \nonumber \\
\omega_1(\lambda)&=& \frac{4}{(2\lambda+1)(2\lambda-1)}.
\ear
In this case, the solution for $\widetilde\omega_2(\lambda)$ is obtained via Fourier transform and can be written as,
\bear
\widetilde\omega_2(\lambda)&=& -\frac{(\lambda^2(4\lambda^2-1)^2 -9)}{27 \lambda^2(4\lambda^2-1)^2} + \frac{c_1}{\sin^2(2\pi\lambda)} + c_2,
\ear
where one has to add some suitable periodic function to ensure the function $\widetilde\omega_2(\lambda)$ is free of pole at $\lambda=0$ which fixes $c_1=-\frac{4}{3}\pi^2$ and it has zero asymptotic which fixes $c_2=\frac{1}{27}$. 

In the homogeneous limit one has,
\bear
\omega_1(0)&=&-4 , \nonumber \\
\widetilde\omega_2(0)&=&\frac{8}{3}-\frac{4}{9}\pi^2= -1.719824178261937\dots , 
\ear
which is in full agreement with the numerical result in Table 1.

Using the previous result, we can also obtain analytically the values of the coefficients $\rho_i(0,0)$ \cite{KNS013},
\bear
\rho_1(0,0)&=\frac{1}{2}-\frac{2}{45}\pi^2&=0.061350915507139\dots , \nonumber\\
\rho_2(0,0)&=-\frac{19}{18}+\frac{14}{135}\pi^2&=-0.032041025072214\dots , \\
\rho_3(0,0)&=-\frac{1}{9}+\frac{4}{135}\pi^2&=0.18132161188412\dots , \nonumber
\ear
which again are in agreement with the numerical results in Table 2. 

It is worth noting that these results for $O(3)$ coincides with the findings for the spin-$1$ $SU(2)$ chain \cite{KNS013}. This is due to the fact its $R$-matrices are exactly same due to the isomorphism between these groups.

\subsubsection{The $O(4)$ case}

Next we show in Table 3 the result for the numerical evaluation of the convolution integral (\ref{gconv}) at the homogeneous point $\lambda=0$ for $n=4$ along with the comparison with the exact diagonalization for finite lattices.
\begin{table}[h]
	\begin{center}
		\begin{tabular}{|l|l|l|}
			\hline  
			Length & $\omega_{1}(0)$  & $\widetilde\omega_{2}(0)$  \\ 
			\hline
			$L=2$ & $-4.000000000000000$ &  $4.000000000000000$ \\
			\hline
			$L=4$ & $-3.000000000000000$ &  $2.250000000000000$ \\
			\hline
			$L=6$ & $-2.868517091819053$      &  $2.057097576517726$ \\
			\hline
			$L\rightarrow\infty$ 
				  & $-2.772588722239781$ &  $1.921812055672715$ \\
			\hline 
		\end{tabular}
	\end{center}
	\caption{Comparison of numerical results from exact diagonalization for $L=2,4,6$ sites with the analytical result in the thermodynamic limit for correlation $\omega_i(0)$ for the $O(4)$ case.}
\end{table}

In addition, we evaluate the coefficients $\rho_i(0,0)$ of the density matrix which are given in Table 4.
\begin{table}[h]
	\begin{center}
		\begin{tabular}{|l|l|l|l|}
			\hline  
			Length & $\rho_{1}(0,0)$  & $\rho_{2}(0,0)$ & $\rho_{3}(0,0)$ \\ 
			\hline
			$L=2$ & $0.000000000000000$ &  $\hspace{0.35cm}0.000000000000000$ & $0.250000000000000$ \\
			\hline
			$L=4$ & $0.034722222222222$ &  $-0.027777777777778$ & $0.138888888888889$  \\
			\hline
			$L=6$ & $0.038254471371995$ &  $-0.029363821569783$ & $0.126345936081805$ \\
			\hline
			$L\rightarrow\infty$ 
			      & $0.040680064040199$ &  $-0.030217991507056$  & $0.117497735346250$  \\
			\hline 
		\end{tabular}
	\end{center}
	\caption{Comparison of numerical results from exact diagonalization for $L=2,4,6$ sites with the analytical result in the thermodynamic limit for the $\rho_i(0,0)$   for the $O(4)$ case.}
\end{table}

Again, the exact diagonalization for finite lattices are in accordance with the infinite lattice result.

Due to the isomorphism $O(4)\sim SU(2)\times SU(2)$, one can write the analytical solution for the case $n=4$ in terms of the solution for the $SU(2)$ spin-$1/2$. The above isomorphism means that the $R$-matrix (\ref{Rmatrix})  for $n=4$ can be decomposed as \cite{MARTINS1997},
\eq
R^{[O(4)]}(\lambda)=R^{[SU(2)]}(\lambda)\otimes R^{[SU(2)]}(\lambda),
\en
where
\eq
R^{[SU(2)]}(\lambda)=\left(
\begin{array}{cccc}
1 & 0 & 0 & 0 \\	
0 & \frac{\lambda}{\lambda+1} & \frac{1}{\lambda+1} & 0 \\	
0 & \frac{1}{\lambda+1} & \frac{\lambda}{\lambda+1} & 0 \\	
0 & 0 & 0 & 1
\end{array}
\right).
\en

This also implies that the two-sites density matrix for $O(4)$ can be written as
\eq
D_2^{[O(4)]}(\lambda_1,\lambda_2)=D_2^{[SU(2)]}(\lambda_1,\lambda_2)\otimes D_2^{[SU(2)]}(\lambda_1,\lambda_2),
\en
where the two-sites density matrix for the $SU(2)$ spin-$1/2$ chain \cite{BOOS05} is given by,  
\eq
D_2^{[SU(2)]}(\lambda_1,\lambda_2)=\left(
\begin{array}{cccc}
	\frac{1}{4}+\frac{\omega(\lambda)}{6} & 0 & 0 & 0 \\	
	0 & \frac{1}{4}-\frac{\omega(\lambda)}{6} & \frac{\omega(\lambda)}{3} & 0 \\	
	0 & \frac{\omega(\lambda)}{3} & \frac{1}{4}-\frac{\omega(\lambda)}{6} & 0 \\	
	0 & 0 & 0 & \frac{1}{4}+\frac{\omega(\lambda)}{6}
\end{array}
\right),
\en
where $\omega(\lambda)=(\lambda^2-1)\frac{d}{d\lambda} \log\left[\frac{\Gamma(1+\frac{\lambda}{2})\Gamma(\frac{1}{2}-\frac{\lambda}{2})}{\Gamma(1-\frac{\lambda}{2})\Gamma(\frac{1}{2}+\frac{\lambda}{2})}\right] +\frac{1}{2}$.

This implies that
\bear
\omega_1(\lambda)&=&\frac{1-2 \omega(\lambda)}{\lambda^2-1}, \\
\widetilde\omega_2(\lambda)&=&\frac{1}{4} \left[\omega_1(\lambda)\right]^2.
\ear

In the homogeneous limit one has,
\bear
\omega_1(0)=&-4 \log2=&2.772588722239781\dots, \nonumber \\
\widetilde\omega_2(0)=&4 (\log2)^2=& 1.921812055672806
\dots , 
\ear
which is in full agreement with the numerical results in Table 3.

Using the previous result, we can also obtain analytically the values of the coefficients $\rho_i(0,0)$,
\bear
\rho_1(0,0)=&\frac{1}{18}+\frac{1}{18}\log2 -\frac{1}{9}(\log2)^2&=0.040680064040196\dots , \nonumber\\
\rho_2(0,0)=&\frac{1}{18}-\frac{5}{18}\log2 +\frac{2}{9}(\log2)^2&=-0.030217991507051\dots , \\
\rho_3(0,0)=&-\frac{1}{36}+\frac{1}{18}\log2 +\frac{2}{9}(\log2)^2&=0.117497735346264\dots , \nonumber
\ear
which again are in agreement with the numerical results in Table 4. 

\section{The $O(5)$ case}

Finally, we show in Table 5 and 6 the result for the numerical evaluation of the convolution integral (\ref{gconv}) at the homogeneous point $\lambda=0$ for $n=5$ along with the comparison with the exact diagonalization results for small lattices. Here the calculation is more subtle since by definition $\widetilde{\omega}_2(0)$ must be zero for non-integers values of ${1}/{\Delta}$. Therefore we had to evaluate its derivative $\widetilde{\omega}^{'}_2(0)$.

\begin{table}[h]
	\begin{center}
		\begin{tabular}{|l|l|l|}
			\hline  
			Length & $\omega_{1}(0)$  & $\widetilde\omega_{2}^{'}(0)$  \\ 
			\hline
			$L=2$ & $-3.333333333333333$ & $6.594547532155967$  \\
			\hline
			$L=4$ & $-2.603912563829967$ & $3.329844337362441$  \\
			\hline
			$L=6$ & $-2.497610064125614$ & $2.955489567911210$  \\
			\hline
			$L\rightarrow\infty$ & $-2.418399152312290$ &  $2.697303095421986$ \\
			\hline 
		\end{tabular}
	\end{center}
	\caption{Comparison of numerical results from exact diagonalization for $L=2,4$ and Lanczos for $L=6$ sites with the analytical result in the thermodynamic limit for correlation $\omega_i(0)$ for the $O(5)$ case.}
\end{table}

\begin{table}[h]
	\begin{center}
		\begin{tabular}{|l|l|l|l|}
			\hline  
			Length & $\rho_{1}(0,0)$  & $\rho_{2}(0,0)$ & $\rho_{3}(0,0)$ \\ 
			\hline
			$L=2$ & $0.000000000000000$ &  \hspace{0.35cm}$0.000000000000000$ & $0.200000000000000$ \\
			\hline
			$L=4$ & $0.024257767401215$ &  $-0.021781358237085$ & $0.100492521231013$  \\
			\hline
			$L=6$ & $0.026877470789173$ &  $-0.023307749344340$ & $0.088920395398474$ \\
			\hline
			$L\rightarrow\infty$ 
			& $0.028642125394723$ & $-0.024107800279869$ & $0.080897173306257$\\			      		      
			\hline 
		\end{tabular}
	\end{center}
	\caption{Comparison of numerical results from exact diagonalization for $L=2,4$ and Lanczos for $L=6$ sites with the analytical result in the thermodynamic limit for the $\rho_i(0,0)$   for the $O(5)$ case.}
\end{table}

Nevertheless, in the $O(5)$ case we could not obtain an analytical solution of the integral (\ref{gconv}), since the inhomogeneous function is given in terms of digamma functions. The analytical evaluation of integrals of this kind has eluded us so far. Integral of similar type also appears in the evaluation of three-sites correlation for the $SU(3)$ integrable spin chain \cite{RIBEIRO2018}, which are also short of analytical solution. Therefore, it is important to obtain analytically the solution of integral of the above kind, since these appear in the evaluation of correlation function of integrable chains of different symmetries.

\section{Conclusion}
\label{CONCLUSION}

We addressed the problem of solving the functional equations of quantum Knizhnik-Zamolodchikov type for the reduced density matrix for integrable $O(n)$ spin chains. 

We have used the functional equation for the $O(n)$ case and other properties like intertwining symmetry, asymptotic, analyticity and normalization to compute the two-sites density matrix elements. 
We give explicit solutions for the two-sites density matrix elements for the $O(n)$ spin chain, which was evaluated for the $n=3,4,\dots,8$ cases at zero temperature.

Although we have shown that the approach of functional equations can be fruitful for explicit computation of correlation functions for $O(n)$ spin chains, we still need to find a closed form for the function $\omega_2(\lambda_1,\lambda_2)$ in order to proceed for correlation at longer distances. 
Another important issue is to investigate the existence factorization of general static correlation functions in
terms of elementary nearest-neighbour function. We expect more naturally that 
the equation (\ref{matrixL-qKZ}) holds for the  $Sp(2m)$ spin chain with the appropriate value of $\Delta=-(m+1)$. It also seems feasible, although more subtle, to extend it for the case of $OSp(n|2m)$ spin chains. We hope to come back to these issues in the future.

\section*{Acknowledgments}

The author thanks the University of Melbourne for support and hospitality and the S\~ao Paulo Research Foundation (FAPESP) for financial support through the grant 2018/25824-0.

\section*{\bf Appendix: Results for $O(6)$, $O(7)$ and $O(8)$}
\setcounter{equation}{0}
\renewcommand{\theequation}{A.\arabic{equation}}
\renewcommand{\baselinestretch}{1.4}

For completeness, in this appendix we present in Tables 7-9 the results for the cases $O(6)$, $O(7)$ and $O(8)$, whose evaluation goes along the same lines as the $O(5)$ case.

\begin{table}[h]
	\begin{center}
		\begin{tabular}{|l|l|l|l|l|l|l|}
			\hline  
			Length & $\omega_{1}(0)$  & $\widetilde\omega_{2}^{'}(0)$ & $\rho_{1}(0,0)$  & $\rho_{2}(0,0)$ & $\rho_{3}(0,0)$ \\ 
			\hline
			$L=2$ & $-3.00000$ &  $3.00000$ & $0.0000000$ &  $\hspace{0.35cm}0.0000000$ & $0.1666667$ \\
			\hline
			$L=4$ & $-2.42539$ &  $1.40209$ & $0.0175797$ &  $-0.0165304$ & $0.0777188$  \\
			\hline
			$L=6$ & $-2.33434$ & $1.19874$  & $0.0197422$ & $-0.0181112$ & $0.0663248$ \\
			\hline
			$L\rightarrow\infty$ 
			   &  $-2.26394$ &  $1.06426$ & $0.0211299$ &  $-0.0188596$  & $0.0587471$  \\		 
			\hline
		\end{tabular}
	\end{center}
	\caption{Results for the $O(6)$ case.}
	\begin{center}
		\begin{tabular}{|l|l|l|l|l|l|l|}
			\hline  
			Length & $\omega_{1}(0)$  & $\widetilde\omega_{2}^{'}(0)$ & $\rho_{1}(0,0)$  & $\rho_{2}(0,0)$ & $\rho_{3}(0,0)$ \\ 
			\hline
			$L=2$ & $-2.80000$ &  $1.879820$ & $0.0000000$ &  $\hspace{0.35cm}0.0000000$ & $0.1428570$ \\
			\hline
			$L=4$ & $-2.32621$ &  $0.831166$ & $0.0132093$ &  $-0.0126998$ & $0.0630920$  \\
			\hline
			$L=6$ & $-2.24665$ &  $0.684007$ & $0.0150284$ & $-0.0142053$ & $0.0518636$ \\
			\hline
			$L\rightarrow\infty$ 
			& $-2.18212$ &  $0.585933$  &  $0.0162103$ & $-0.0149653$ & $0.0443502$ \\
			\hline
		\end{tabular}
	\end{center}
	\caption{Results for the $O(7)$ case.}
	\begin{center}
		\begin{tabular}{|l|l|l|l|l|l|l|}
			\hline  
			Length & $\omega_{1}(0)$  & $\widetilde\omega_{2}^{'}(0)$ & $\rho_{1}(0,0)$  & $\rho_{2}(0,0)$ & $\rho_{3}(0,0)$ \\ 
			\hline
			$L=2$ & $-2.66667$ &  $1.347915$ & $0.0000000$ &  $\hspace{0.35cm}0.0000000$ & $0.1250000$ \\
			\hline
			$L=4$ & $-2.26376$ &  $0.572129$ & $0.0102433$ &  $-0.0099689$ & $0.0530226$  \\
			\hline
			$L=6$ & $-2.19338$ &  $0.454788$ & $0.0117758$ & $-0.0113250$ & $0.0421188$ \\
			\hline
			$L\rightarrow\infty$ 			
			& $-2.13339$ &  $0.374522$  &  $0.0128029$ & $-0.0120623$ & $0.0346391$ \\
			\hline
		\end{tabular}
	\end{center}
	\caption{Results for the $O(8)$ case.}
\end{table}

\renewcommand{\baselinestretch}{1.5}

\end{document}